\documentclass[a4paper,12pt]{article}

\usepackage{epsfig}
\usepackage{amssymb}

\def\bc{\begin{center}}
\def\ec{\end{center}}

\def\be{\begin{equation}}
\def\ee{\end{equation}}
\def\beq{\begin{eqnarray}}
\def\eeq{\end{eqnarray}}
\def\bfig{\begin{figure}}
\def\efig{\end{figure}}
\def\bnum{\begin{enumerate}}
\def\enum{\end{enumerate}}

\begin{document}

\begin{flushright}
Journal-Ref: Astronomy Letters, 2007, v. 33, No 9, pp. 594-603
\end{flushright}
 
\vspace{1cm}

\bc {\LARGE\bf Hydrodynamic Processes in Young Binary Systems as a
Source of Cyclic Variations of Circumstellar Extinction}\\
\vspace{0.4cm}
{\bf N.Ya.~Sotnikova$^1$\footnotetext[1]{email: nsot@astro.spbu.ru},
V.P.~Grinin$^{1,2}$}\\
\vspace{0.4cm}
1 -- {\it Sobolev Astronomical Institute, St. Petersburg State University,
Universitetskii pr. 28, \\
Petrodvorets, St. Petersburg, 198504 Russia,}\\
2 -- {\it Pulkovo Astronomical Observatory, Russian Academy of Sciences,
Pulkovskoe sh. 65, St. Petersburg, 196140 Russia}
\ec

%\vspace{0.5cm}
%=============================================================================
\bc
{\bf Abstract}
\ec
%=============================================================================

Hydrodynamic models of a young binary system accreting matter from the
remnants of a protostellar cloud have been calculated by the SPH
method. It is shown that periodic variations in column density in
projection onto the primary component take place at low
inclinations of the binary plane to the line of sight. They can
result in periodic extinction variations. Three periodic
components can exist in general case. The first component has a
period equal to the orbital one and is attributable to the streams
of matter penetrating into the inner regions of the binary. The
second component has a period that is a factor of 5-8 longer than
the orbital one and is related to the density waves generated in a
circumbinary (CB) disk. The third, longest period is attributable
to the precession of the inner asymmetric region of CB disk. The
relationship between the amplitudes of these cycles depends on the
model parameters as well as on the inclination and orientation of
the binary in space. We show that at a dust-to-gas ratio of $1 :
100$ and and a mass extinction coefficient of 250~cm$^2$~g$^{-1}$,
the amplitude of the  brightness variations of the primary
component in the V-band can reach $1^m$ at a mass accretion rate
onto the binary components of $10^{-8}\,M_{\odot}$ yr$^{-1}$ and a
$10^{\rm o}$ inclination of the binary plane to the line of sight. We
discuss possible applications of the model to pre-main-sequence
stars.


\vspace{1cm}
%=============================================================================
\bc
INTRODUCTION
\ec
%=============================================================================

Binarity is widespread among stars, including young
pre-main-sequence stars (see the review by Mathieu et al. 2000).
Such stars still continue accreting matter from the remnants of a
protostellar cloud. Numerical simulations by Artymowicz and Lubow
(1994, 1996) of hydrodynamic processes in young binaries show that
periodic gravitational perturbations and viscous forces produce a
matter-free gap at the binary center into which two streams of
matter generally unequal in intensity from a circumbinary (CB)
disk penetrate. These streams maintain the accretion activity of
the binary components.

The simulations by Artymowicz and Lubow (1996) showed that the accretion
rate in binaries with eccentric orbits depends on the orbital phase,
reaching its maximum at the time of periastron passage. For this reason,
in binaries whose components are cool young T Tauri stars (their luminosity
depends sensitively on the accretion rate), periodic brightness variations
in their components can take place. This prediction of the theory was
confirmed by observations (Mathieu et al. 2000).

In this paper, we reproduce the Artymowicz--Lubow model by
the SPH method and show that another type of cyclic photometric
variability of a young binary due to periodic extinction
variations on the line of sight is possible in this model. In
contrast to periodic modulation of the accretion rate, periodic
extinction variations can be observed in binaries with both
elliptical and circular orbits, provided they are inclined at a small
angle to the line of sight.

%=============================================================================
\bc
FORMULATION OF THE PROBLEM
\ec
%=============================================================================

Following Artymowicz and Lubow (1994, 1996), we assume that the
circumbinary disk is coplanar with the binary. As in the papers of
these authors, the hydrodynamic calculations presented below were
performed by the smoothed particle hydrodynamics (SPH) method
(Lucy 1977; Gingold and Monagan 1977) according to a scheme
similar to that suggested by Hernquist and Katz (1989), but with a
constant smoothing length of hydrodynamic quantities. No thermal
balance was calculated. The system was assumed to be isothermal.
The SPH implementation used was described in detail previously
(Sotnikova 1996).

%=============================================================================
\bc
{\it The Hydrodynamic Model}
\ec
%=============================================================================

The mass of the gaseous disk was assumed to be negligible compared to the
total mass of the stars in the binary. This allowed the disk self-gravity to
be neglected. In the initial scheme, we assumed free boundary conditions,
i.e., neglected the pressure at the boundary of the gas distribution.
This approximation is justified if a flow of cold gas is modeled. During
our calculations, it emerged that the presence of a free boundary led to a
slow outward expansion of the disk, which deteriorated the statistics when
the temporal variations in disk column density were determined. Therefore,
the scheme was slightly modified: an artificial potential barrier on the
far periphery of the disk was superimposed on the general gravitational
field of the
binary\footnote{Note that a similar barrier was used with the same goal
by Artymowicz and Lubow (1994, 1996).}.
The barrier parameters were chosen in such a way that, on the one hand,
the disk dissipation was slowed down and, on the other hand, no significant
distortions were introduced into the dynamics of density waves in the inner
disk regions. Test calculations showed that introducing a barrier reduced
the dissipation rate of the CB disk approximately twofold. In the models
presented below, the dissipation is attributable mainly to mass accretion
onto the binary components: through this process, the number of particles
in the CB disk decreases approximately twofold after 600 binary revolutions.
In the case without a barrier, the same result is obtained after 300
revolutions.

The SPH equations of motion for particle $i$ that represents an element of 
gas in are very similar to those described in (Hernquist and Katz~1989)
$$
\frac{d {\bf v}_i}{d t} = - m \sum_j\left(
\frac{2\,\sqrt{P_i \, P_j}}{\rho_i \rho_j} + Q_{ij}\right)
\nabla_i W({\bf r}_i - {\bf r}_j;h) - \nabla \varphi({\bf r}_i) \, ,
$$
where $P_i$, $P_j$ , $\rho_i$, $\rho_j$ are the pressure and density at
the position ${\bf r}_i$ and ${\bf r}_j$ of particles $i$ and $j$;
for the isothermal case under consideration, the equation of state is
$P = c^2 \rho$, where $c$ is the speed of sound; $\varphi$ is a variable
gravitationalpotential produced by the binary; $m$ is the mass of the SPH
particles (we considered equal-mass particles); $W$ is the kernel for
smoothing hydrodynamicquantities (it was chosen in the form of a spline;
seeMonaghan and Gingold 1983); $h$ is the smoothing length.

The contribution from the artificial viscosity to the pressure gradient is
described by the tensor $Q_{ij}$. There are various representations of
$Q_{ij}$. We used its expression suggested by Monaghan and Gingold (1983).
For advantages and disadvantages of this choice, see Hernquist and Katz (1989).

As in the papers by Artymowicz and Lubow, we took into account the
contribution from the viscous terms in the cases where the SPH particles
approached each other and recede from each other, i.e., in the form
$$
Q_{ij} = (-\alpha c \mu_{ij} + \beta \mu_{ij}^2) / \rho_{ij} \, ,
$$
where $\mu_{ij} =
h\,({\bf v}_i-{\bf v}_j) \cdot ({\bf r}_i-{\bf r}_j)/(r_{ij}^2 + \eta^2)$,
$\rho_{ij}=(\rho_i+\rho_j)/2$, $r_{ij} = |{\bf r}_i-{\bf r}_j|$,
$\eta \simeq 0.1\,h$.
The parameters $\alpha$ and $\beta$ are analogues of the viscosity
coefficients in the Navier-Stokes equation. Following Artymowicz and Lubow
(1994), we assumed for most of the models that $\alpha \simeq 1$ and
$\beta = 0$.

The choice of parameter $c$, an isothermal speed of sound, is critical for
our models. It defines the effective viscosity of the gaseous disk:
$\nu \sim \alpha c h$. Following Artymowicz and Lubow (1994), we chose this
parameter in units of the velocity of a test particle in a circular orbit with
a radius equal to the semimajor axis of the binary a moving around a point
mass $m_1 + m_2$, where $m_1$ and $m_2$ are the masses of the binary
components. The parameter $c$ in these units was varied in the range from
0.01 to 0.08. Reducing the viscous properties of the disk reduced the
contribution from hydrodynamic effects and the behavior of the system
was similar to that of a celestial-mechanical system. Below, we call gaseous
disks with $c \approx 0.01-0.02$ and $c = 0.05$ ``cold'' and ``warm''
CB disks, respectively.

The smoothing length was fixed at $h = 0.1 a$, where $a$ is the orbital
semimajor axis of the secondary component. This allowed the hydrodynamic
quantities to be smoothed over 40-60 neighboring particles at a typical number
of particles $N \sim 60\,000$.

To integrate the SPH equations, we used the standard explicit leapfrog scheme;
the time step $dt$ was controlled by the Courant condition.

%=============================================================================
\bc
{\it Initial Conditions and Binary Parameters}
\ec
%=============================================================================

The number of test particles modeling the CB disk was chosen to be from
50\,000 to 75\,000. The particles were distributed in accordance with a
surface density profile $\sim 1/r$. The radius of the matter-free gap at the
initial time was taken to be $r_{\rm in} = 2a$. After several binary
revolutions, it virtually ceased to change and did not differ much from its
initial value. We put the outer boundary of the disk at the initial time at a
distance $r_{\rm out} = 5.8 a$. Thus, we modeled more extended disks
than Artymowicz and Lubow~(1994). The periodic variations in gravitational
potential produced no perturbations in this region of the gaseous
disk and the existence of a boundary (in particular, the presence of a
barrier) had no effect on the inner disk regions. The vertical particle
distribution followed a barometric law.

At the initial time, the particles were placed in circular orbits around the
center of mass of the binary with a Keplerian velocity corresponding to the
orbital radius.

We varied the orbital eccentricity of the binary within the range from
$e = 0$ to $e = 0.7$ and the component mass ratio $q = m_2 / m_1$ within
the range from 0.1 to 1.0. The evolution of the CB disk was traced on time
scales up to 300 binary revolutions (in some cases, up to 600 revolutions).
The orbital parameters of the binary in this time interval were assumed to be
constant.

%=============================================================================
\bc {\it Determining the Column Density in the Disk} \ec
%=============================================================================

Simultaneously with the calculation of the dynamical evolution of the gaseous
disk in the binary's periodically varying potential, we determined the mass
accretion rate from the disk onto both components. We assumed that if a
particle fell into a region less than 0.3 of the radius of the corresponding
Roche lobe, then it was captured by the star and contributed to the accretion
rate. Subsequently, such particles were eliminated from our calculations.
The derived accretion rates of test particles onto the binary components
were then used to determine the particle mass when calculating the
circumstellar extinction.

The particle column density was determined for various orbital phases and
various inclinations of the line of sight to the binary plane. Let us denote
the binary inclination to the line of sight by $\theta$. To calculate the
particle column density $n(\theta, \, t)$ as a function of time, we chose a
column with a cross section $\sigma = 0.1 a \times 0.2 a$. Test calculations
showed that the statistical fluctuations due to a small number of test
particles in the column increase at lower values of $\sigma$, while
the features on the $t$ dependences of $n$ are smoothed at higher values
of $\sigma$.

%=============================================================================
\bc
RESULTS OF SIMULATIONS
\ec
%=============================================================================

As an example, Fig.~1 shows the particle distribution in the binary after
60 revolutions from the beginning of our calculation. The models with warm
(a) and cold (b) '' disks are shown. The model parameters are:
$e = 0.5$, $m_2 : m_1 = 0.7 : 2$, $c = 0.02$ for the cold disk; $c = 0.05$
for the warm disk.

\begin{figure}[!hb]
\centerline{\psfig{file=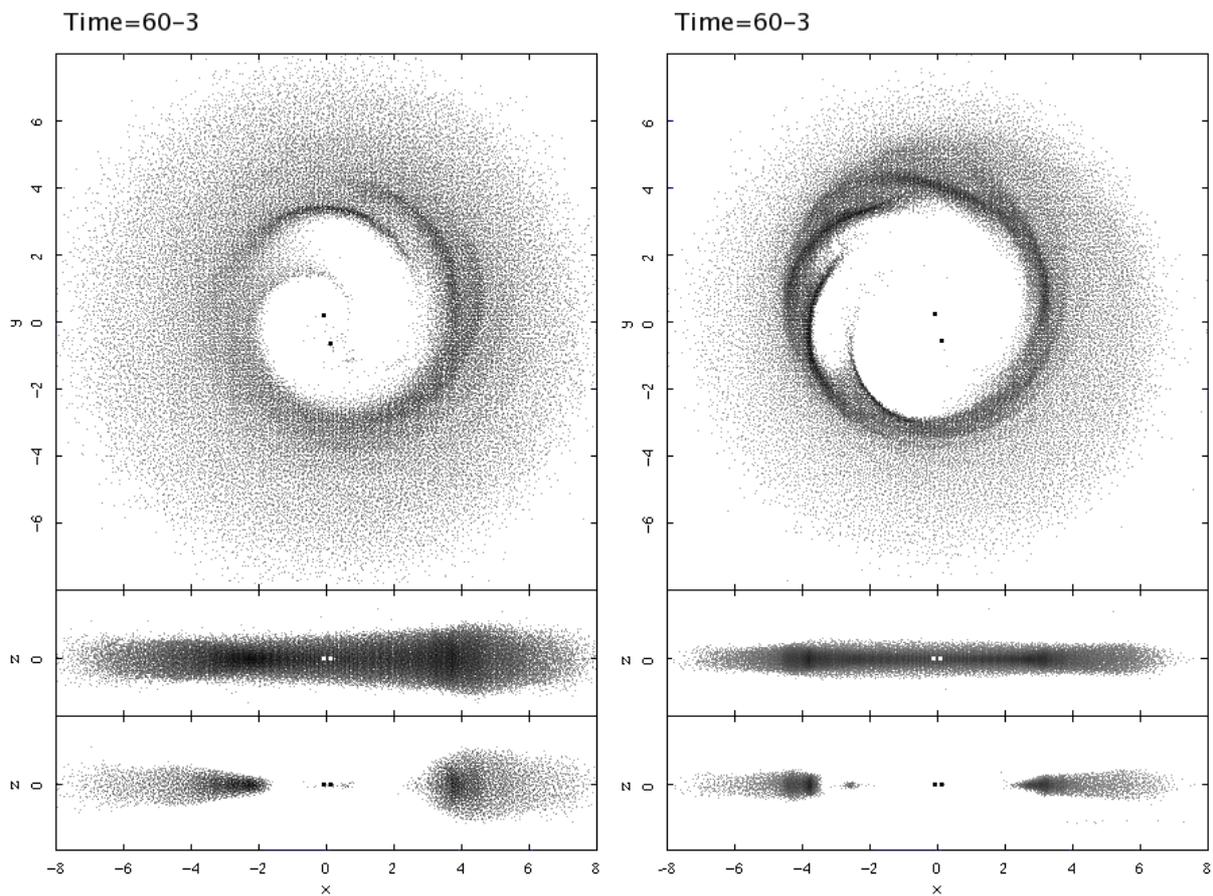,width=16cm}}
\caption[1]{Binary models with warm (a) and cold (b) '' disks and a low-mass
($m_2 : m_1 = 0.7 : 2$) secondary component in an eccentric ($e = 0.5$) orbit.
The figure shows the following: view from the pole (upper panels), projection
onto the $xz$ plane (middle panels), and section (lower panels) passing
through the center of mass in this plane. The binary is displayed in
revolution 61 at phase 3/8.
The strenght of the image blackening corresponds
to the logarithm of the particle number per pixel. }
\label{fig1}
\end{figure}

In the model with a warm disk (Fig.~1a), we clearly see two streams of matter
unequal in intensity from the CB disk, which feed the accretion disks of the
binary components and which are extensions of the spiral density waves
(the second stream is more pronounced at other intermediate orbital phases).
The more intense stream accretes onto the less massive binary component. This
peculiarity was first pointed out by Artymowicz and Lubow (1996) and, as the
calculations by Bate and Bonnell (1997) showed, is obtained even in the
models with a component mass ratio of $1 : 10$. This is because the low-mass
companion moves in its orbit not far from the inner boundary of the CB
disk --- the main reservoir of the matter that it ``pulls'' on itself, while
the primary component of the binary is located near its center, in a
matter-free zone. The characteristic size of this zone depends on orbital
eccentricity $e$ and component mass ratio $q$ and is equal to
$\approx (2-3) a$ (Artymowicz and Lubow (1994). On the whole, the disk more
likely resembles a wide ring.

In the ``cold'' model (Fig.~1b), the spiral pattern on the CB disk appears
more fragmentary and is represented by several short remnants of the spiral
density waves near the inner disk boundary. The accretion streams are less
pronounced. The disk itself is geometrically thinner.

%----------------------------------------------------
\bc {\it The Global Asymmetry of CB disk} \ec
%----------------------------------------------------
%
We see from Figs.~1a~and~1b that the distribution of matter in the CB disk is
characterized by a global asymmetry seen in both projections of the binary.
The asymmetry manifests itself particularly clearly in the inner disk
region --- the inner boundary of the ring has a noticeable eccentricity and
its center is shifted relative to the binary center of mass. According to
Lubow and Artymowicz (2000), such an eccentricity is the result of
instability. It arises when there is a $3 : 1$ resonance in the gaseous
disk. This instability manifests itself even in binaries with circular orbits.
For this instability to operate, the mass ratio of components $q$ must be
greater than $0.2$. On the other hand, the component
masses must differ, $q$ cannot be close to unity, or the instability will be
damped by viscosity.

Additional effects arise in the case of eccentric orbits. The presence of an 
one-armed bar potential with $(m,\, l) = (1,\, 0)$ results in lopsided
structure of a disk. The disk disturbance produced by a variable
gravitational potential follows the apsidal motion of the binary. The disk
eccentricity varies periodically and the rate of these variations depends on
the angle between the major axes of the binary orbit and the elliptical disk
itself. The effects of disk precession due to the quadrupole moment from the
binary are added to this. The precession rate is low and the entire eccentric
disk turns in a time interval that is hundreds of times longer than the
orbital period.

If we look at the disk edge-on, then its significant asymmetry is also
noticeable in the vertical direction: on the one side, it is considerably
thicker than on the other side (Fig.~1, central panels). We are probably
the first to point out this peculiarity of CB disks. The part of the disk
whose inner boundary is located farther from the center of mass turns out to
be thicker. The thickening may be caused by a weakening of the gravitational
forces from the binary, which is known to result in an increase in the
geometrical thickness of the disk in an external gravitational field.

Figure~2 demonstrates the vertical asymmetry and the overall turn of the
warm disk on time scales of the order of one precession period
($\approx 200\,P$, where $P$ is the orbital period of the binary). The
parameters of the presented model are $e = 0.5$, $m_2 : m_1 = 1 : 2$, and
$c = 0.05$. As we will show below, the existence of a global asymmetry in the
CB disk is one of the reasons why the optical properties of a young binary
at low inclinations to the line of sight depend significantly on its
orientation in space.

\begin{figure}[!hb]
\centerline{\psfig{file=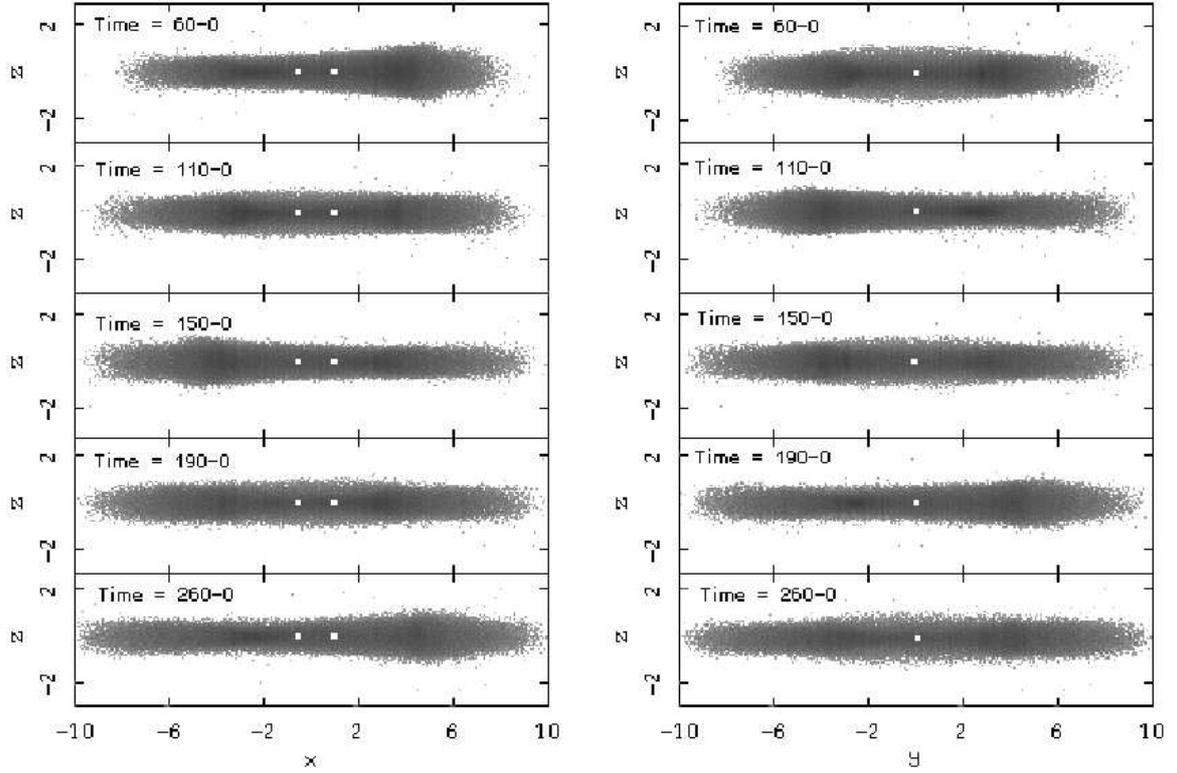,width=16cm}}
\caption[2]{Precession of the warm disk in projections onto the $xz$ and
$yz$ planes at various times. The time is given in units of the orbital period.
The model parameters are $m_2 : m_1 = 1 : 2$ and $e = 0.5$.
The strenght of the image blackening corresponds
to the logarithm of the particle number per pixel. }
\label{fig2}
\end{figure}

The global asymmetry in the cold disk on time scales of several hundred binary
revolutions is more pronounced than that in the warm disk. This is
particularly clearly seen if we look at the disk edge-on. Its precession time
is approximately twice that for the warm disk. Figure~3 shows the projections
of the cold disk onto the $xz$ and $yz$ planes at various times. In a time
of about 200 orbital periods, the disk in the $yz$ projection turns
approximately through $180^{\rm o}$, while in the ``warm'' model the CB disk
made an almost complete turn. The CB disk precession and asymmetry, which is
particularly clearly seen in the vertical direction, are common properties of
such disks and take place not only in the models with eccentric orbits, but
also in those with circular orbits. Moreover, we found no significant
dependence of the precession period on eccentricity.

\begin{figure}[!hb]
\centerline{\psfig{file=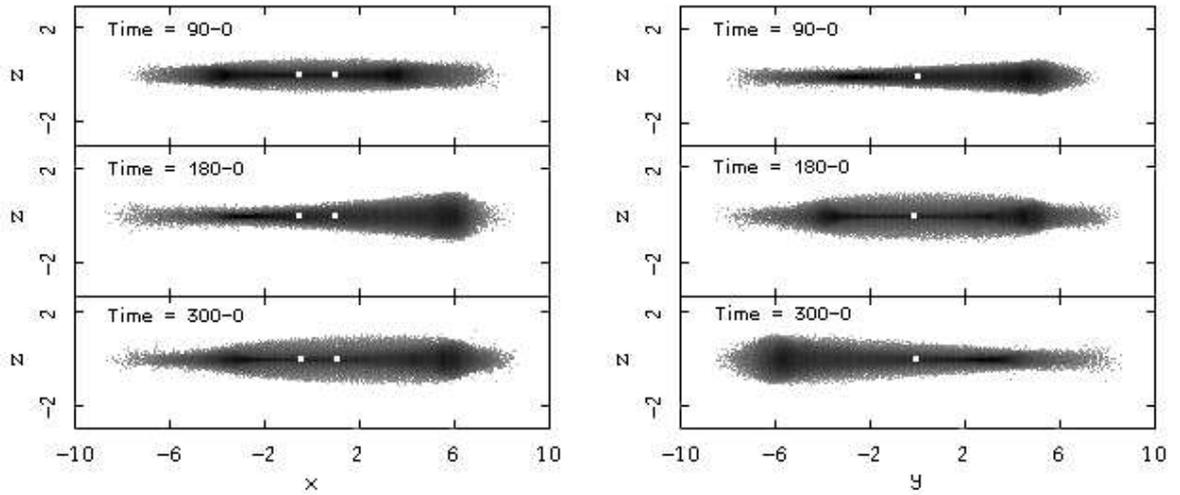,width=16cm}}
\caption[3]{
Same as Figure~\ref{fig2} for the cold disk. The binary parameters are
$m_2 : m_1 = 0.7 : 2$ and $e = 0.5$. The strenght of the image blackening
corresponds to the logarithm of the particle number per pixel. }
\label{fig3}
\end{figure}

%=============================================================================
\bc
{\it The Behavior of the Column Density}
\ec
%=============================================================================

Our calculations showed that the particle column density in binaries with
elliptical orbits toward the primary component depends not only on the
orbital phase and the orbital inclination to the line of sight $\theta$,
but also on the orientation of the orbit relative to the observer. Figure~4
shows the behavior of the column density $n(\theta,\, t)$ for two angles,
$\theta = 0^{\rm o}$ and $\theta = 10^{\rm o}$, and one of the orbital
orientations relative to the observer at which the azimuth angle between the
direction from the primary component to the apoastron of the secondary star
and the observer's direction is zero. In this case, the orbital apoastron of
the secondary component lies between the observer and the primary component.

\begin{figure}[!hb]
\centerline{\psfig{file=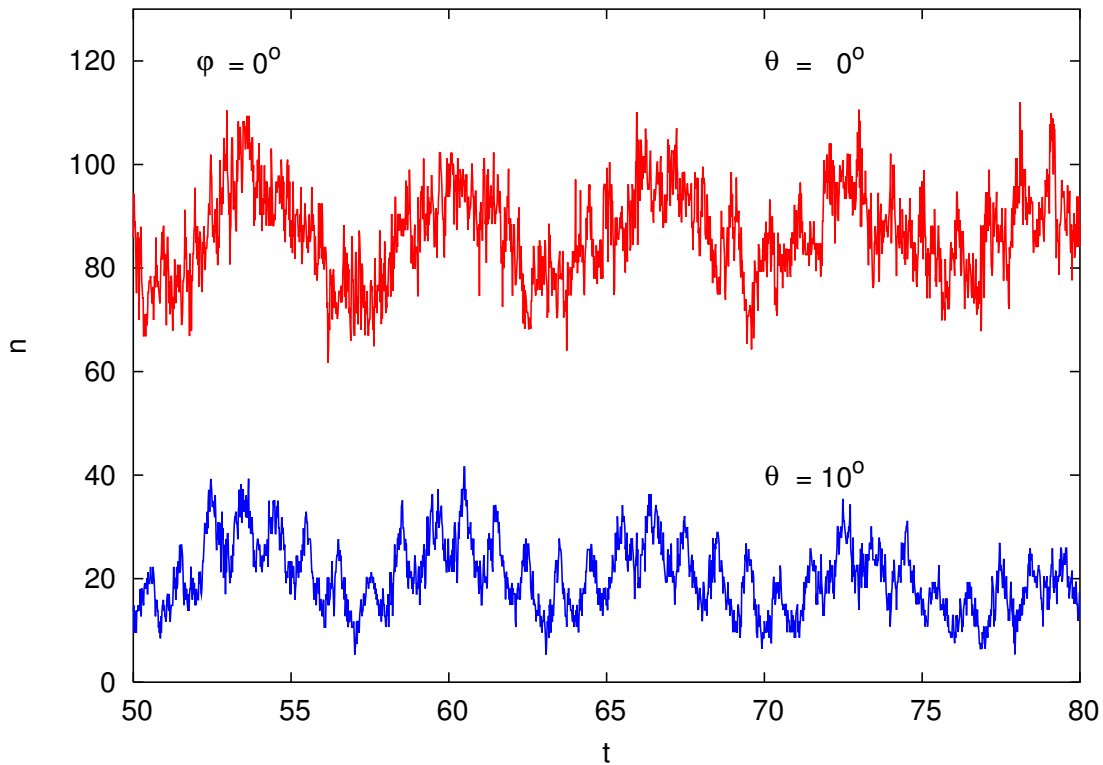,angle=-90,width=15cm}}
\caption[4]{Variations in column density $n(\theta,\, t)$ for two orbital
inclinations with respect to the observer: $\theta = 0^{\rm o}$
(solid line) and $\theta = 10^{\rm o}$ (dotted line). The orbit is turned to
the observer in such a way that the apoastron of the secondary component lies
between the observer and the primary star. The model parameters are
$m_2 : m_1 = 1 : 2$, $e = 0.5$, the warm '' disk. The time is given in units
of the orbital period. }
\label{fig4}
\end{figure}

We see from Fig.~4 that two cycles are present in the column density
variations. One of them has a period equal to the orbital period. Its
amplitude is larger in the case with an inclination of $10^{\rm o}$ and is
barely noticeable in the case with $\theta = 0^{\rm o}$. This cycle originates
from the streams of matter accreting onto the binary components and
periodically crossing the line of sight. The second component in the
dependence on model parameters has a period that is approximately a factor
of 6-8 longer than the orbital one. It owes its origin to the motion of
density waves in the inner CB disk region. Comparison of the solutions
obtained at the same $\theta$, but at different binary orientations relative
to the observer shows that the relationship between these two components
depends sensitively on the observer's direction. This dependence originates
from the global CB disk asymmetry discussed above. The slow secular turn of
an asymmetric disk changes the relationship between the amplitudes of the
short and longer periods of the column density variations at the same orbital
inclination.

\begin{figure}[hb]
\centerline{\psfig{file=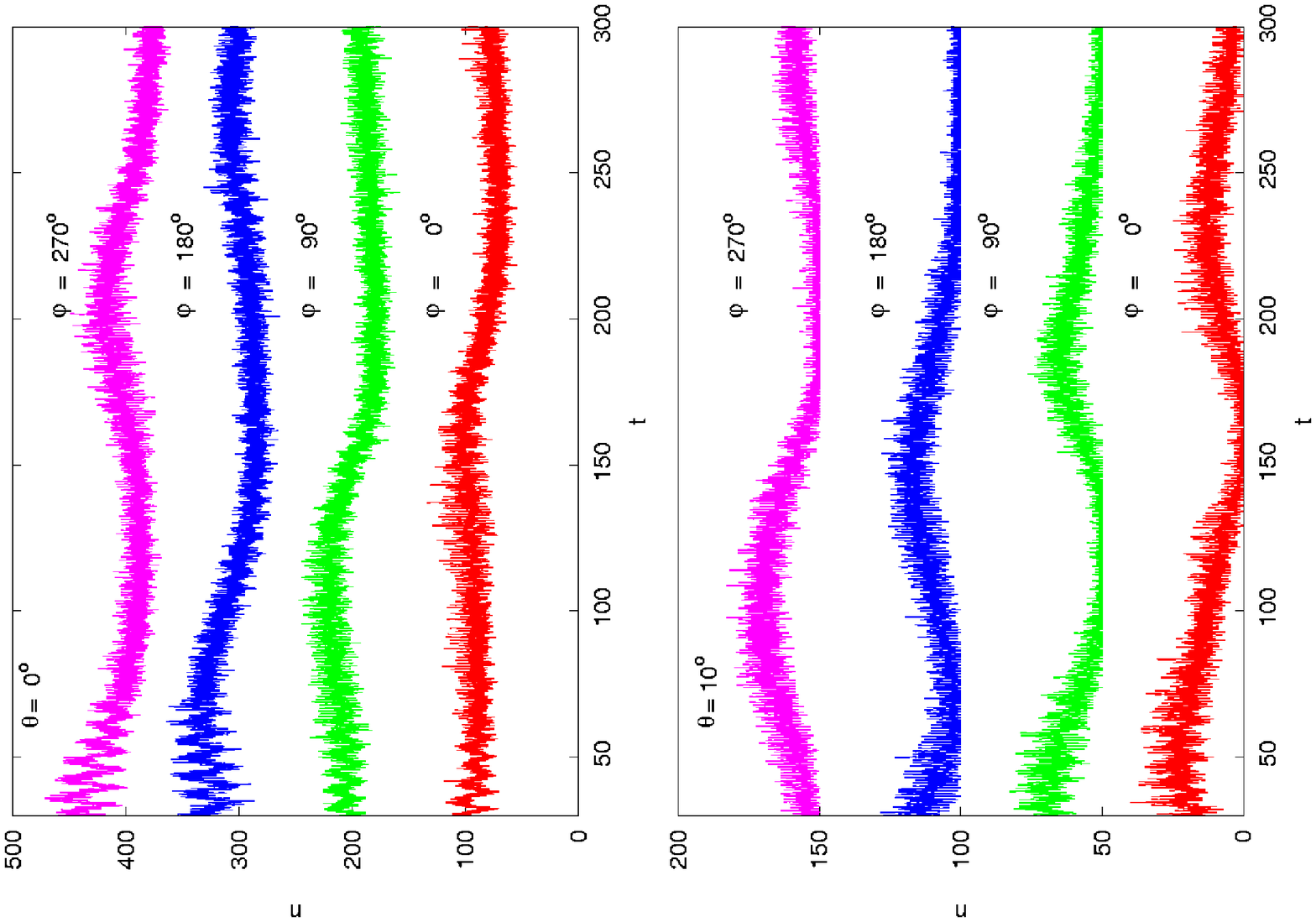,angle=-90,width=14cm}}
\caption[5]{Variations in column density $n(\theta,\, t)$ for two orbital
inclinations with respect to the observer: (a) $\theta = 0^{\rm o}$ and
(b) $\theta = 10^{\rm o}$. The variations in column density with time are
shown in each plot for one of the four orbital orientations with respect to
the observer (from the bottom upward: the angles $\phi$ between the directions
of the apoastron of the secondary component and the observer as viewed from
the primary star are, respectively, $0^{\rm o}$, $90^{\rm o}$, $180^{\rm o}$,
and $270^{\rm o}$). For convenience, the upper and lower plots were displaced
with respect to one another by 100 and 50 units along the $y$ axis,
respectively. The model parameters are $m_2 : m_1 = 0.7 : 2$ and $e = 0.5$,
the warm CB disk. The time is given in units of the orbital period. }
\label{fig5}
\end{figure}

Figures~5~and~6 show the column density variations on a long time scale
(300 orbital periods) for four orbital orientations relative to the observer
corresponding to azimuth angles $\phi$ equal to $0^{\rm o}$, $90^{\rm o}$,
$180^{\rm o}$, and $270^{\rm o}$ (the angles are measured from the direction
of the apoastron of the secondary in the direction opposite to its orbital
motion). As has already been noted above, in the first of these cases, the
orbital apoastron of the secondary component lies between the observer and the
primary component. In the third case, the binary is viewed from the periastron.
In addition, for each of the four listed cases, we considered two orbital
inclinations: $\theta = 0^{\rm o}$ (Figs.~5a~and~6a) and $\theta = 10^{\rm o}$
(Figs.~5b~and~6b).

Slow CB disk precession produces a long-period modulation of $n(\theta,\, t)$.
As a result, apart from the two periodic components discussed above, a secular
period is also present in the column density variations with time. There is
also a slow decrease in column density with time common to all models that
is attributable to gradual CB disk dissipation due to mass accretion onto the
binary components. The secular period of the column density variations is
equal to the precession period of the inner region of an asymmetric CB disk
and is $\sim 200\,P$ for the models presented in Fig.~2. In observations at
$\theta = 0^{\rm o}$, the line of sight passes through different parts of the
asymmetric gaseous ring. As a result, the values of $n$ in the global
modulation curve for the same time, but for different orbital orientations
relative to the observer are different (Figs.~5a~and~6a). In the directions
(along the line of sight) where the disk is more extended, the column density
is higher.

In observations at $\theta = 10^{\rm o}$, the picture is significantly
different (Figs.~5b~and~6b). Over fairly long time intervals, the column
density decreases almost to zero. At these times, the thin side of the disk
is turned to the observer and the line of sight hardly touches it. The small
$n$ variations in these time intervals stem from the fact that the line of
sight periodically crosses the stream of matter directed to the secondary
component. Thus, because of the CB disk precession, a young binary whose
equatorial plane is slightly inclined to the line of sight can be observable
in certain time intervals and may turn out to be completely occulted from the
observer by its own CB disk after a lapse of time and this occultation can
last very long.

\begin{figure}[!hb]
\centerline{\psfig{file=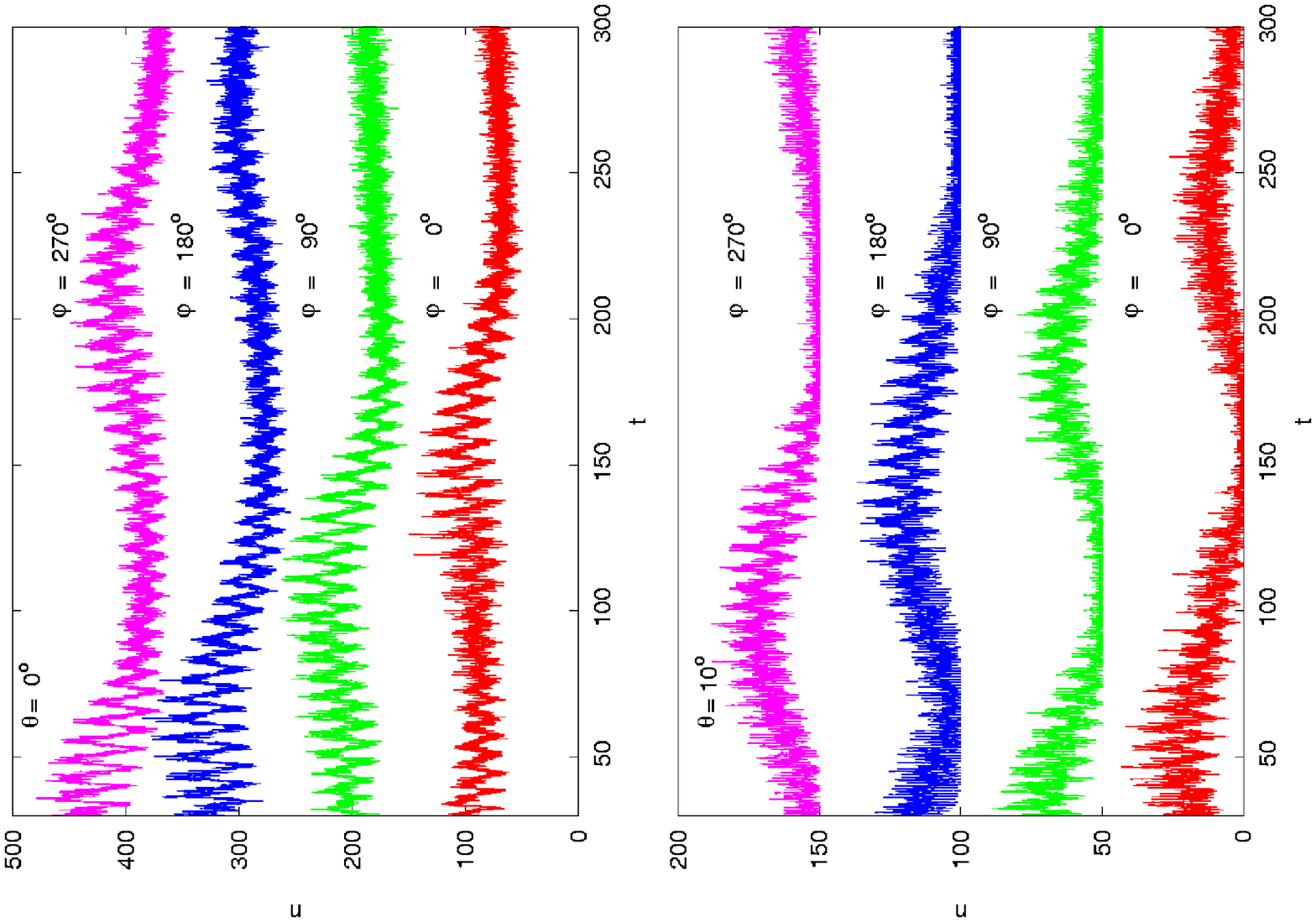,angle=-90,width=14cm}}
\caption[6]{
Same as Figure~\ref{fig5} for $m_2 : m_ 1 = 1 : 2$. }
\label{hernq-fig}
\end{figure}

In general case, the three different (in duration) periods of the
column density variations manifest themselves as follows. The
secular (precession) period manifests itself clearly in the models
with both circular and eccentric orbits. A decrease in viscosity
causes an increase in precession period (for the models with $e =
0.5$ and $m_2 : m_1 = 0.7 : 2$, the secular period for the cold
disk increases by almost a factor of 2 compared to the warm model
and is $\sim 400\, P$). For binaries with high eccentricities ($e
= 0.5, 0.7$), the period related to the motion of density waves at
the inner disk boundary (5-8 orbital periods, depending on the
model parameters) is superimposed on the secular period. This
second period shows up particularly clearly at the maxima of the
secular $n(\theta,\, t)$ modulation amplitudes (Figs.~5~and~6). In
this case, its amplitude in ``cold'' binaries is smaller than that
in ``warm'' ones.

Since apart from gas, there is also dust in the streams of matter penetrating
from the CB disk into the inner region of the binary, the periodic variations
in column density will be accompanied by periodic extinction variations.
Taking the standard dust-to-gas ratio for the model considered above to be
$1 : 100$ and the mass extinction coefficient to be 250 cm$^2$ g$^{-1}$, we
estimated the photometric effect due to the periodic column density
variations. The amplitude of the V-band brightness variations in the primary
component at a $10^{\rm o}$ inclination of the binary plane to the line of
sight turned out to be about 1$^m$ at an accretion rate of
$10^{-8}\,M_{\odot}$ yr$^{-1}$. In binaries with a higher accretion rate, an
appreciable (in amplitude) cyclic activity can be observed at a higher binary
inclination to the line of sight.

%=============================================================================
\bc
CONCLUSIONS
\ec
%=============================================================================

The results of our hydrodynamic calculations presented above show that cyclic
variations in the column density of matter accreting onto the binary
components can take place in young binaries inclined at a small angle to the
line of sight. The fundamental period of these variations is equal to the
orbital one and is attributable to the streams of matter in the inner parts
of the binary that are periodically projected onto the primary component. The
second period is produced by the motion of spiral density waves in the CB
disk and is a factor of 5-8 longer than the orbital one. At low inclinations
of the binary, both $n(\theta,\, t)$ oscillation modes can be present
simultaneously. The relationship between them depends both on the inclination
of the binary plane and on its orientation in space.

Apart from these two cycles, there is also a secular cycle with a duration of
the order of several hundred orbital periods in the behavior of the column
density. This cycle is attributable to the precession of the inner parts of
an asymmetric CB disk. The lower the viscosity, the longer its duration. The
characteristics of the two shorter cycles (primarily their amplitude) depend
significantly on the phase of the secular cycle (Figs.~5~and~6). Note also
that the existence of a global asymmetry in the distribution of matter in the
inner CB disk region must be taken into account when the intrinsic
polarization in young binary stars is modeled.

The periodic variations in column density can be accompanied by noticeable
extinction and brightness variations in the binary. Such variations are
observed in UX Ori stars (Shevchenko et al. 1993; Grinin et al. 1998;
Rostopchina et al. 2000; Bertout 2000). The brightness of these stars is
known to undergo great variations attributable to extinction variations in
circumstellar disks inclined at a small angle to the line of sight (see the
review by Grinin (2000) and references therein). Both ``rapid'' variability
on time scales of the order of several days with an irregular (unpredictable)
pattern and slow variability on a time scale from several to twenty years
or more are observed. In a number of stars, the slow component is cyclic in
pattern. In two cases, SV Cep (Rostopchina et al. 2000) and CQ Tau
(Shakhovskoi et al. 2005), the cyclic activity is described by two
oscillation modes with a period ratio close to that obtained above (5-8).
Therefore, the idea that the cyclic activity of UX Ori stars can be the result
of their latent binarity seems quite plausible.

Other objects of application of the theory considered above can be
such young binaries with abnormally long occultations as KH 15D
(Winn et al. 2006), H 187 (Grinin et al. 2006; Nordhagen et al.
2006), and GW Ori (Shevchenko et al. 1998). There are the reasons
to assume that the equatorial planes of these binaries are
inclined at a small angle to the line of sight and, hence,
periodic variations in column density can take place. We are going
to return to a discussion of these questions in the next papers.

%=============================================================================
\bc
ACKNOWLEDGMENTS
\ec
%=============================================================================

We thank Pawel Artymowicz for a helpful discussion of the
questions touched upon in the paper. This work was performed as
part of the ``Origin and Evolution of Stars and Galaxies'' Program
of the Presidium of the Russian Academy of Sciences under support
of INTAS grant no. 03-51-6311 and grant no. NSh-8542.2006.2.

%=============================================================================
\bc
REFERENCES
\ec
%=============================================================================

1. P. Artymowicz and S. H. Lubow, Astrophys. J. {\bf 421}, 651 (1994).

2. P. Artymowicz and S. H. Lubow, Astrophys. J. {\bf 467}, L77 (1996).

3. M. R. Bate and I. A. Bonnell,Mon. Not. R. Astron. Soc. {\bf 285}, 33 (1997).

4. C. Bertout, Astron. Astrophys. {\bf 363}, 984 (2000).

5. R. A. Gingold and J. J. Monagan, Mon. Not. R. Astron. Soc. {\bf 181}, 375 (1977).

6. V. P. Grinin, Astron. Soc. Pac. Conf. Ser. {\bf 219}, 216 (2000).

7. V. P. Grinin, O. Yu. Barsunova, S. G. Sergeev, et al., Pis'ma Astron. Zh.
{\bf 32}, 918 (2006) [Astron. Lett. {\bf 32}, 827 (2006)].

8. V. P. Grinin, A. N. Rostopchina, and D. N. Shakhovskoi,
Astron. Lett. {\bf 24}, 802 (1998).

9. L. Hernquist and N. Katz, Astrophys. J., Suppl. Ser. {\bf 70}, 419 (1989).

10. S. H. Lubow and P. Artymowicz, Protostars and Planets IV, Ed. by
V. Mannings, A. Boss, and S. S. Russell (Univ. Arizona Press, Tucson, 2000),
p.731.

11. L. B. Lucy, Astron. J. {\bf 82}, 1013 (1977).

12. R. D. Mathieu, F. C. Adams, and D. W. Latham, Astron. J. {\bf 101}, 2184 (1991).

13. R. D. Mathieu, A. M. Ghez, E. K. N. Jensen, and M. Simon,
Protostars and Planets IV, Ed. by V. Mannings, A. Boss, and S. S. Russell
(Univ. Arizona Press, Tucson, 2000), p. 559.

14. J. J. Monaghan and R. A. Gingold, J. Comp. Phys. {\bf 52}, 374 (1983).

15. S. Nordhagen, B. Herbst, E. C. Williams, and E. Semkov,
astro-ph/0606501 (2006).

16. A. N. Rostopchina, V. P. Grinin, D. N. Shakhovskoi, et al.,
Astron. Zh. {\bf 77}, 420 (2000) [Astron. Rep. {\bf 44}, 365 (2000)].

17. D. N. Shakhovskoi, V. P. Grinin, and A. N. Rostopchina,
Astrofzika {\bf 48}, 165  (2005) [Astrophys. {\bf 48}, 135 (2005)].

18. V. S. Shevchenko, K. N. Grankin,M. A. Ibragimov, et al.,
Astrophys. Space Sci. {\bf 202}, 121 (1993).

19. V. S. Shevchenko, K. N. Grankin, S. Yu. Mel'nikov, and S. A. Lamzin,
Pis'ma Astron. Zh. {\bf 24}, 614 (1998) [Astron. Lett. {\bf 24}, 528 (1998)].

20. N. Ya. Sotnikova, Astrofzika {\bf 39}, 259 (1996) [Astrophys. {\bf 39}, 141 (1996)].

21. J. N. Winn, C. M. Hamilton, W. J. Herbst, et al.,
Astrophys. J. {\bf 644}, 510 (2006).

\begin{flushright}
\it Translated by V. Astakhov
\end{flushright}

\end{document}